\documentclass[journal]{IEEEtran}

\usepackage[T1]{fontenc}
\usepackage{graphicx}
\usepackage[caption=false]{subfig}
\usepackage{mathrsfs,amsmath,amsfonts,amssymb,amsthm,amstext,amscd,amsxtra,amsopn,mathtools}
\usepackage{diagbox}
\usepackage{lettrine}
\usepackage{cite}
\usepackage{eucal}
\usepackage{soul}
\providecommand{\keywords}[1]{\textbf{\textit{Index terms---}} #1}
\usepackage{bm}
\usepackage{multirow}
\usepackage{esint}
\usepackage{empheq}
\usepackage{comment}
\usepackage{boxhandler}
\usepackage{bm}
\usepackage{algorithm}
\usepackage[noend]{algpseudocode}
\usepackage{etoolbox}
\apptocmd{\sloppy}{\hbadness 10000\relax}{}{}
\usepackage{leftidx}
\usepackage{makecell}
\usepackage{stackrel}
\usepackage{xcolor}
\usepackage{hyphenat}
\usepackage{nicefrac}
\usepackage{siunitx}
\usepackage[normalem]{ulem}
\usepackage[hyphens]{url}
\usepackage[hidelinks]{hyperref}
\hypersetup{breaklinks=true}
\renewcommand{\vec}[1]{\bm{#1}}

\newcommand{\matvec}[1]{\mathbf{#1}}

\algnewcommand{\IfThenElse}[3]{
  \State \algorithmicif\ #1\ \algorithmicthen\ #2\ \algorithmicelse\ #3}
\algnewcommand{\IfThen}[3]{
  \State \algorithmicif\ #1\ \algorithmicthen\ #2}

\DeclarePairedDelimiter\abs{\lvert}{\rvert}

\newcommand\Aop{{\CMcal{A}}}

\newcommand\Gop{{\CMcal{G}}}

\newcommand\Zop{{\CMcal{Z}}}

\newcommand\At{{\cal{A}}}

\newcommand\Mt{{\cal{M}}}
\newcommand\Nt{{\cal{N}}}

\makeatletter
\def\BState{\State\hskip-\ALG@thistlm}

\makeatother

\algnewcommand{\Initialize}[1]{%
  \State \textbf{Initialize:}
  \Statex \hspace*{\algorithmicindent}\parbox[t]{.8\linewidth}{\raggedright #1}
}

\DeclarePairedDelimiter\ceil{\lceil}{\rceil}

\newcommand{\iu}{{\mathrm{i}}}

\let\oldref\ref
\renewcommand{\ref}[1]{(\oldref{#1})}

\makeatletter
\DeclareRobustCommand{\pder}[1]{%
  \@ifnextchar\bgroup{\@pder{#1}}{\@pder{}{#1}}}
\newcommand{\@pder}[2]{\frac{\partial#1}{\partial#2}}
\makeatother

\newcolumntype{M}[1]{>{\hbox to #1\bgroup\hss$}l<{$\egroup}}

\makeatletter
\newcommand\@brcolwidth{0.67em}

\def\@brarray[#1]{\array{r*\c@MaxMatrixCols {M{#1}}}}
\makeatother

\title{A hybrid volume\hyp surface integral equation method for rapid electromagnetic simulations in MRI}

\author{Ilias I. Giannakopoulos, Georgy D. Guryev, Jos{\'e} E. C. Serrall{\'e}s, Jan Pa\v{s}ka, Bei Zhang, \\ Luca Daniel, ~\IEEEmembership{Fellow,~IEEE}, Jacob K. White, ~\IEEEmembership{Fellow,~IEEE}, Christopher M. Collins, ~\IEEEmembership{Senior Member,~IEEE} \\ Riccardo Lattanzi, ~\IEEEmembership{Senior Member,~IEEE}

\thanks{This work was supported in part by research grants from the National Institutes of Health (NIH P41 EB017183, and R01 EB024536) and the National Science Foundation (NSF CAREER 1453675) in the United States, and was performed under the rubric of the Center for Advanced Imaging Innovation and Research (CAI$^2$R, www.cai2r.net). \textit{(Corresponding author: Ilias I. Giannakopoulos.)}}
\thanks{Ilias I. Giannakopoulos, Jan Pa\v{s}ka, Christopher M. Collins and Riccardo Lattanzi are with the Center for Advanced Imaging Innovation and Research (CAI$^2$R), and with the Bernard and Irene Schwartz Center for Biomedical Imaging, Department of Radiology, New York University Grossman School of Medicine, NY 10016 USA (e-mail: ilias.giannakopoulos@nyulangone.org).}
\thanks{Bei Zhang is with the Advanced Imaging Research Center, UT Southwestern Medical Center, 2201 Inwood Rd, Dallas, TX 75390 USA.}
\thanks{Georgy D. Guryev, Jos{\'e} E. C. Serrall{\'e}s, Luca Daniel, and Jacob K. White are with the Research Laboratory of Electronics, Department of Electrical Engineering and Computer Science, Massachusetts Institute of Technology, Cambridge, MA 02139 USA.}}

\begin{document}
\bstctlcite{IEEEexample:BSTcontrol}

\maketitle

\begin{abstract}
\textit{Objective:} We developed a hybrid volume surface integral equation (VSIE) method based on domain decomposition to perform fast and accurate magnetic resonance imaging (MRI) simulations that include both remote and local conductive elements. \textit{Methods:} We separated the conductive surfaces present in MRI setups into two domains and optimized electromagnetic (EM) modeling for each case. Specifically, interactions between the body and EM waves originating from local radiofrequency (RF) coils were modeled with the precorrected fast Fourier transform, whereas the interactions with remote conductive surfaces (RF shield, scanner bore) were modeled with a novel cross tensor train\hyp based algorithm. We compared the hybrid\hyp VSIE with other VSIE methods for realistic MRI simulation setups. \textit{Results:} The hybrid\hyp VSIE was the only practical method for simulation using 1 mm voxel isotropic resolution (VIR). For 2 mm VIR, our method could be solved at least 23 times faster and required 760 times lower memory than traditional VSIE methods. \textit{Conclusion:} The hybrid\hyp VSIE demonstrated a marked improvement in terms of convergence times of the numerical EM simulation compared to traditional approaches in multiple realistic MRI scenarios. \textit{Significance:} The efficiency of the novel hybrid\hyp VSIE method could enable rapid simulations of complex and comprehensive MRI setups. 
\end{abstract}


\section*{Nomenclature}
\begin{IEEEdescription}[\IEEEsetlabelwidth{Notation }]
\item[Notation] Description      
\item[$a$] Scalar                                 	           
\item[$\vec{a}$] Vector in $\mathbb{C}^3$                 		   
\item[$\matvec{a}$] Vector in $\mathbb{C}^{n}$                          
\item[$A$] Matrix in $\mathbb{C}^{n_1 \times n_2}$             
\item[$A^{\rm T}$] Transpose of matrix                                
\item[$A^*$] Conjugate transpose of matrix                       
\item[$\bar{A}$] Conjugate of matrix                                 
\item[$\Aop$] Tensor in $\mathbb{C}^{n_1 \times n_2 \times n_3 \times \dots}$  
\item[$\At$] Operator acting on vectors in $\mathbb{C}^3$      
\item[$\mathfrak{G}$] Dyad          
\item[$\iu$] Imaginary unit $\iu^2 = -1$ 
\end{IEEEdescription}

\section{Introduction} \label{sc:I}
\IEEEPARstart{A}{ccurate} electromagnetic (EM) modeling of radiofrequency (RF) coils and their interaction with biological tissue is critical in magnetic resonance (MR) imaging (MRI) \cite{ibrahim2005design}. For example, due to the high prototyping cost of multi\hyp element coil arrays \cite{duyn2012future, fujita2013rf, hernandez2020review, tavaf2021self}, it is desirable to optimize coil design in simulation before construction. Coil optimization is particularly important at ultra\hyp high\hyp field (UHF) MRI, since poorly designed RF coils could worsen image quality and compromise patient safety, due to the RF field inhomogeneities and spatially non\hyp uniform specific absorption rate amplifications \cite{jin1997sar, lattanzi2009electrodynamic, zhang2013complex, kraff2015mri, erturk2017toward, ladd2018pros}.
\par
Although MRI simulation extensions have been incorporated into commercial EM simulation software, these tools could have high memory requirements and long solution times for problems involving multi-channel coils, because they are based on finite difference and finite element approaches. The computational burden further increases for UHF MRI because for accurate simulations one also needs to include the RF shield, due to its effect on the EM field and signal-to-noise ratio (SNR) distributions \cite{zhang2021effect}. The MAgnetic Resonance Integral Equation (MARIE) suite \cite{villenamarie} was developed to address the limitations of general-purpose commercial software. MARIE is an open-source software specifically tailored to model EM interactions between biological tissue and RF coils at MR frequencies. MARIE employs the volume\hyp surface integral equation (VSIE) method, which is especially useful for single\hyp frequency problems like MRI, since the arising linear system of equations can be easily adapted to existing iterative numerical linear algebra routines, and solved rapidly. Algorithms implemented on top of MARIE have demonstrated improved convergence times \cite{Polimeridis2014, Villena2016, giannakopoulos2019memory} and superior accuracy \cite{tambova2017generalization, georgakis2020fast} compared with conventional methods. However, the solution time in MARIE can still considerably increase at UHF when using fine voxel resolutions and including an RF shield for accurate coil simulations.
\par
To address this, one could take advantage of the low\hyp rank properties of the VSIE coupling matrix \cite{Chai2013}. For example, an algorithm based on the precorrected fast Fourier transform (pFFT) was presented in \cite{guryev2019fast, milshteyn2021individualized} to accelerate MARIE in cases where the conductors and the object are close to each other. Techniques for matrix compression have instead been proposed to improve simulation efficiency in the case of conductors placed away from the sample (e.g., RF shield) \cite{giannakopoulos2021compression}. In this work, we developed a hybrid\hyp VSIE method for MARIE to rapidly simulate realistic experimental setups, with conductors positioned at various distances from the sample. The proposed method assembles a low-parametric representation of the VSIE by combining multiple approaches: pFFT \cite{guryev2019fast}, adaptive cross approximation (ACA) \cite{Bebendorf2003}, Tucker decomposition \cite{giannakopoulos2019memory}, and tensor train (TT)\hyp cross \cite{oseledets2010tt}. As a result, the memory footprint is considerably reduced and simulations can be rapidly executed on a graphical processing unit (GPU) even for fine voxel resolutions.
\par
The remainder of this article is organized as follows. In Section II, we review the technical background on surface and volume integral equations methods, as well as TT and Tucker decomposition methods. In section III we introduce our proposed hybrid\hyp VSIE. Section IV and Section V describe the numerical experiments and associated results, respectively. A discussion of the results is provided in Section VI, whereas Section VII summarizes the main conclusions of this work.

\section{Technical Background}
\subsection{Integral Equation Method in MRI}
The EM phenomena in MR can be thoroughly explained with Maxwell's equations \cite{maxwell1865viii}. The single operating frequency of the MR scanner allows for the development of fast and customizable algorithms for the solution of the integral equation form of Maxwell's equations. In such a representation, the coil and the sample are replaced with equivalent surface and volumetric current sources \cite{chew2008integral, Oijala2014} that radiate the same electromagnetic waves as the coil and the body, respectively: 
\begin{equation}
\begin{aligned}
\vec{j}_{\rm c}\!\left(\vec{r} \right) &= \hat{n}\!\left(\vec{r} \right) \times \vec{h}_{\rm c}\!\left(\vec{r} \right) \\
\vec{j}_{\rm b}\!\left(\vec{r} \right) &= \iu \omega \epsilon_0 \chi_e\!\left(\vec{r} \right) \vec{e}_{\rm b}\!\left(\vec{r} \right).
\end{aligned}
\end{equation} 
Here, $\vec{j}_{\rm c}\!\left(\vec{r} \right)$ and $\vec{j}_{\rm b}\!\left(\vec{r} \right)$ are the equivalent surface and polarization body currents, respectively, $\vec{r}$ is the observation position vector, $\hat{n}\!\left(\vec{r} \right) \times \vec{h}_{\rm c}\!\left(\vec{r} \right)$ is the magnetic field impressed on the coil's surface, $f = \omega / (2\pi)$ is the Larmor frequency of interest, and $\epsilon_0$ is the vacuum permittivity. $\chi_e\!\left(\vec{r} \right) = \epsilon_r\!\left(\vec{r} \right)-1$, $\epsilon_r\!\left(\vec{r} \right)$, and $\vec{e}_{\rm b}\!\left(\vec{r} \right)$ are the electric susceptibility, relative permittivity, and the total electric field of the object, respectively. 
\par
The surface integral equation (SIE) used to compute $\vec{j}_{\rm c}\!\left(\vec{r} \right)$ is the electric field integral equation \cite{Rao1982}:
\begin{equation}
\frac{\iu \hat{n}\!\left(\vec{r} \right)}{\omega \mu_0} \times \vec{e}^{\rm inc}_{\rm c}\!\left(\vec{r} \right) = \hat{n}\!\left(\vec{r} \right) \times \iint\limits_{\rm c} \mathfrak{G}\!\left(\vec{r},\vec{r}'\right) \vec{j}_{\rm c}\!\left(\vec{r}'\right) d^2\vec{r}',
\label{eq:sie}
\end{equation}
where $\rm c$ refers to the coil's surface, $\hat{n}\!\left(\vec{r} \right)$ is the normal vector on $\rm c$, $\mu_0$ is the vacuum permeability, and $\vec{r}'$ is the source position vector. $\vec{e}^{\rm inc}_{\rm c}\!\left(\vec{r} \right)$ is the incident electric field on $\rm c$ and could be interpreted as a voltage excitation at a one of the feeding ports of the coil, as in \cite{jiao1999fast, lo2011finite}. $\mathfrak{G}(\vec{r},\vec{r}')$ is the dyadic homogeneous Green's function \cite{tai1994dyadic}:
\begin{equation}
\begin{aligned}
\mathfrak{G}(\vec{r},\vec{r}') &= \frac{1}{4\pi} \!\left(\mathfrak{I} + \frac{\nabla \nabla}{k_0^2} \right)g(\vec{r},\vec{r}') \\
g(\vec{r},\vec{r}') &= \frac{e^{-\iu k_0 \abs{\vec{r}-\vec{r}'}}}{4 \pi \abs{\vec{r}-\vec{r}'}},
\end{aligned}
\end{equation}
where $\mathfrak{I}$ is the unit dyad, $k_0$ is the vacuum wave number, and $g(\vec{r},\vec{r}')$ is the free\hyp space Green's function.
\par
The volume integral equation (VIE) used to compute $\vec{j}_{\rm b}\!\left(\vec{r} \right)$ is the current\hyp based volume integral equation \cite{Polimeridis2014}:
\begin{equation}
\!\left(\Mt_{\epsilon_r\!\left(\vec{r} \right)} - \Mt_{\chi_e\!\left(\vec{r} \right)} \Nt \right) \vec{j}_{\rm b}\!\left(\vec{r} \right) = \iu \omega \epsilon_0 \Mt_{\chi_e\!\left(\vec{r} \right)} \vec{e}^{\rm inc}_{\rm b}\!\left(\vec{r} \right).
\label{eq:vie}
\end{equation}
Here $\rm b$ refers to the object (``body''), $\Mt$ is a multiplication operator that multiplies quantities with the parameter indicated in subscript, $\vec{e}^{\rm inc}_{\rm b}\!\left(\vec{r} \right)$ is the incident electric field on $\rm b$, and $\Nt$ is the following Green's function integro\hyp differential operator.
\begin{equation}
\Nt \!\left(\vec{j}_{\rm b}\!\left(\vec{r} \right)\right) \triangleq \nabla \times \nabla \times \iiint\limits_{\rm b} g\!\left(\vec{r},\vec{r}'\right) \vec{j}_{\rm b}\!\left(\vec{r}'\right) d^3\vec{r}'.
\label{eq:N}
\end{equation}

When the coil is loaded with a body, the incident electric field on the body $\vec{e}^{\rm inc}_{\rm b}\!\left(\vec{r} \right)$ is the electric field scattered from the coil to the body, which is directly related to the current distribution on the coil $\vec{j}_{\rm c}\!\left(\vec{r} \right)$. Since such current distribution is perturbed by the back\hyp scattered field from the body, it cannot be simply computed by solving equation \eqref{eq:sie}, as in the unloaded case. To account for the back\hyp scattering effect due to the EM interactions between coil and body, one needs to solve \eqref{eq:sie} and \eqref{eq:vie} in tandem, by constructing a coupled VSIE.

\subsection{Discretized VSIE Linear System}
SIEs and VIEs are usually solved with the Galerkin Method of Moments (MoM) \cite{harrington1993field}, where the coils and the body are discretized. In this work, we followed the techniques presented in \cite{Villena2016} and built our method on top of the open\hyp source software MARIE \cite{villenamarie}. Next, we briefly review the construction of the VSIE discretization system of linear equations. 
\par
Coils are modeled with a triangular tessellation and the unknown $\vec{j}_c$ is approximated with the Rao\hyp Wilton\hyp Glisson (RWG) basis functions \cite{Rao1982}. After applying the Galerkin method, \eqref{eq:sie} becomes the linear system 
\begin{equation}
Z_{\rm cc} \matvec{j}_{\rm c} = \matvec{v},
\label{eq:disc_sie}
\end{equation}
where $Z_{\rm cc} \in \mathbb{C}^{m \times m}$ is the discretized counterpart of the integral in the right\hyp hand\hyp side of \eqref{eq:sie}, where the coil is discretized into $m$ elements. $\matvec{v} \in \mathbb{C}^{m \times 1}$ and $\matvec{j}_{\rm c} \in \mathbb{C}^{m \times 1}$ represent the voltage excitation vector and the unknown coil currents in each of the $m$ discretization elements of the coil. For coarse resolutions of the coil mesh, multiplications involving $Z_{\rm cc}$ are fast, because the matrix is small and dense. When the coil's mesh resolution is fine, or when additional conductive structures, such as the scanner's bore, or an RF shield, are included in the simulation, the size of the $Z_{\rm cc}$ matrix can become large. In such cases, since the off-diagonal blocks are low\hyp rank \cite{wei2012fast}, one can compress them with a matrix decomposition method, for example, a pseudo\hyp skeleton approximation \cite{goreinov1997theory, goreinov1995pseudo, Goreinov2001, Kurz2002} or ACA \cite{Bebendorf2003, Zhao2005}, to reduce $Z_{\rm cc}$'s memory demands. 
\par
The discretization of the VIE in \eqref{eq:vie} also results in a linear system:
\begin{equation}
Z_{\rm bb} \matvec{j}_{\rm b} = \matvec{e}.
\label{eq:disc_vie}
\end{equation}
Here, $Z_{\rm bb} \in \mathbb{C}^{q \cdot n_{\rm v} \times q \cdot n_{\rm v}}$ is the discretized version of the left\hyp hand\hyp side of \eqref{eq:vie}, $\matvec{j}_{\rm b} \in \mathbb{C}^{q \cdot n_{\rm v} \times 1}$ are the unknown body currents in each of the $n_{\rm v}$ discretization elements of the body, for each of the $q$ components of the basis used to approximate the polarization currents, and $\matvec{e} \in \mathbb{C}^{q \cdot n_{\rm v} \times 1}$ is the discretized external incident field that illuminates the body. To approximate $\vec{j}_{\rm b}\!\left(\vec{r} \right)$ in each voxel, one can employ polynomial basis functions, either piecewise constant (PWC) \cite{Polimeridis2014}, which use $q = 3$ unknowns per discretization element, or piecewise linear (PWL) \cite{georgakis2020fast}, for which $q = 12$. In either case, since the body is a 3D object, the matrix $Z_{\rm bb}$ is usually too large to be fully assembled. One can address this by exploiting the translational invariance property of $g\!\left(\vec{r},\vec{r}'\right)$ to enforce a multilevel block\hyp Toeplitz structure on $Z_{\rm bb}$. Precisely, if we enclose the body in a uniform voxelized domain of size $n_1 \times n_2 \times n_3 = n_{\rm v}$, then only the first columns of each of the Toeplitz blocks of $Z_{\rm bb}$ need to be assembled and the computations required for the EM simulations can be rapidly executed using the fast Fourier transform (FFT) \cite{Catedra1989, Zwamborn1992}. In addition, these defining columns of $Z_{\rm bb}$ can be reshaped into $q \times q$ 3D tensors, each of dimensions $n_1 \times n_2 \times n_3$, which can be vastly compressed with the Tucker decomposition \cite{giannakopoulos2019memory}.
\par
Since conductive surfaces (e.g., RF coils) and body are both present in the MRI simulation setup, equations \eqref{eq:disc_sie}, \eqref{eq:disc_vie} must be solved together through a coupled block system of equations:
\begin{equation}
\begin{bmatrix}
Z_{\rm cc} & \!\left(Z_{\rm cb}\right)^{\rm T} \\
Z_{\rm cb} & Z_{\rm bb}        
\end{bmatrix} \begin{bmatrix}
\matvec{j}_{\rm c} \\
\matvec{j}_{\rm b} 
\end{bmatrix} = \begin{bmatrix}
\matvec{v} \\
0
\end{bmatrix}.
\label{eq:nvsie}
\end{equation} 
Note that $\matvec{e}$ from \eqref{eq:disc_vie} was discarded, since it can be expressed as the product between the new matrix $Z_{\rm cb} \in \mathbb{C}^{q \cdot n_{\rm v} \times m}$ and $\matvec{j}_{\rm c}$. $Z_{\rm cb}$ is called the \textit{coupling matrix} and maps the equivalent surface electric currents on the conductors onto the electric fields that illuminate the body, via the dyadic Green's function. In contrast with $Z_{\rm cc}$ and $Z_{\rm bb}$, the coupling matrix is not the outcome of a Galerkin inner product, but of a Petrov\hyp Galerkin one, because the testing and basis functions used for its construction are different \cite{Oijala2014}.

\subsection{Compression of the Coupling Matrix}
One can reshape the coupling matrix to a higher\hyp order tensor based on the dimensions of the simulation domain. Specifically, each column of the coupling matrix is readily convertible to a 4D tensor of dimensions $n_1 \times n_2 \times n_3 \times q$, while the whole coupling matrix can be interpreted as a 5D tensor with dimensions $n_1 \times n_2 \times n_3 \times m \times q$. In this way, the matrix can be compressed with a tensor decomposition\hyp based algorithm of choice and large\hyp scale EM simulations can be practically performed \cite{giannakopoulos2021compression}.
\par
The $5$\hyp th dimension of the reshaped coupling matrix corresponds to the number of basis functions per voxel and is not compressible for the functions of choice (PWC and PWL). However, $q$ is small comparing to the remaining dimensions, therefore the 5D tensor can be treated as $q$ independent 4D tensors with low\hyp rank properties \cite{Chai2013}. To exploit the low\hyp rankness and compress the coupling matrix, one can consider two cases: \textit{a)} Conductors at a distance from the body, such as in the case of the scanner bore, the scanner body coil \cite{vaughan2004efficient, milshteyn2021individualized} or other RF shields \cite{zhang2021effect}, and \textit{b)} conductors placed close to the body, such as in the case of local receive coils \cite{corea2016screen, tavaf2021self}. 
\par
For case \textit{a)}, a 4D tensor decomposition method can be employed for the compression of the coupling matrix \cite{giannakopoulos2021compression, giannakopoulos2021atensor}. For case \textit{b)}, the $4$\hyp th dimension $m$ cannot be compressed as much as the first three dimensions, so a 4D tensor decomposition becomes suboptimal. One can instead use $m$ independent 3D tensor decompositions \cite{giannakopoulos2021compression}, which, however, yields slow matrix\hyp vector product times for fine coil discretizations (high values of $m$). Alternatively, it is possible to define an extended voxelized domain of dimensions $n^{\rm ext}_1 \times n^{\rm ext}_2 \times n^{\rm ext}_3$, with $n^{\rm ext}_i \geq n_i, \: i=1,2,3$, which encloses both coil and body, and then project each triangular pair of the coil mesh onto an expansion domain of voxels \cite{phillips1997precorrected}. Then, the VSIE problem could be transformed into an equivalent VIE problem using the pFFT method as in \cite{guryev2019fast}. The resulting extended VIE domain is a uniform voxelized grid, like the original VIE one, so the Green's function operators can be compressed with the methods presented in \cite{giannakopoulos2019memory}. 

\subsection{Tensor Decomposition Methods}
Tensor decomposition methods are able to provide a low\hyp parametric representation of specific $d$\hyp dimensional arrays that possess low\hyp rank properties. Available methods include the canonical polyadic model \cite{carroll1970analysis, harshman1970foundations}, the Tucker decomposition \cite{Tucker1966, Oseledets2008}, the tensor train \cite{oseledets2010tt, Oseledets2011}, the hierarchical Tucker decomposition \cite{ballani2013black}, and the tensor ring decomposition \cite{zhao2016tensor}, with various applications in electrodynamics \cite{chen2019sparsity, wang2020voxcap, qian2021compression} and MRI \cite{yaman2019low, yu2014multidimensional}. Next we briefly review TT and Tucker decomposition since they are used in this work. 

\subsubsection{Tensor Train (TT)}
A $d$ dimensional tensor $\Aop \in \mathbb{C}^{n_1 \times n_2 \times \cdots \times n_d}$ can  be approximated by a tensor $\tilde{\Aop}$ of the same dimensions through TT as:
\begin{equation}
\tilde{\Aop}_{i_1 i_2 \cdots i_d} = \hspace{-1.5em} \sum\limits_{j_1,j_2,\cdots,j_{d-1}}^{r_1,r_2,\cdots,r_{d-1}} \hspace{-1.5em} G^1_{i_1 j_1} \Gop^2_{j_1 i_2 j_2} \cdots \Gop^{d-1}_{j_{d-2}i_{d-1} j_{d-1}} G^d_{j_{d-1} i_d}.
\label{eq:TT}
\end{equation}
Here $r_k$, for $k = 1,\dots,d-1$ are the TT\hyp ranks and the matrices $G^1 \in \mathbb{C}^{n_1 \times r_1}$, $G^d \in \mathbb{C}^{r_{d-1} \times n_d}$ and tensors $\Gop^k \in \mathbb{C}^{r_{k-1} \times n_k \times r_k}$, $k = 2,\dots,d-1$ are the \textit{cores} or \textit{carriages} of the TT. 
\par
For any tensor $\Aop$ with low TT\hyp rank, one can always find a low\hyp rank quasi\hyp optimal TT approximation with a prescribed accuracy, using a sequence of SVDs on the unfolding matrices of $\Aop$ \cite{Oseledets2011}. One could also use a TT\hyp cross approximation method \cite{oseledets2010tt, savostyanov2011fast} to construct the TT of $\Aop$. In this case, the tensor $\Aop$ has to be replaced with a function that maps from indices in $\Aop$ to corresponding values in $\Aop$. Then, the TT\hyp cross method, iteratively calls this function as a black\hyp box procedure in order to construct the TT form of $\Aop$ \cite{oseledets2010tt, ballani2013black}.
\par
When the TT ranks are small, TT\hyp cross can be considerably faster than TT\hyp SVD, because the former does not require assembly of all elements of $\Aop$. This can be particularly useful when compressing the $Z_{\rm cb}^{\rm f}$ coupling matrix, because each of its elements is a 5D surface\hyp volume integral, which can be costly for a large number of quadrature integration points. 

\subsubsection{Tucker Decomposition}
Tucker decomposition approximates the previously mentioned tensor $\Aop$ as:
\begin{equation}
\tilde{\Aop}_{i_1 i_2 \cdots i_d} \approx \sum\limits_{j_1,j_2,\cdots,j_d}^{r_1,r_2,\cdots,r_d} \Gop_{j_1 j_2 \cdots j_d} U^1_{i_1 j_1} U^2_{i_2 j_2} \cdots U^d_{i_d j_d}.
\end{equation}
Here, $\Gop \in \mathbb{C}^{r_1 \times r_2 \times \cdots \times r_d}$ is the \textit{Tucker core}, $U^k \in \mathbb{C}^{n_k \times r_k}$, $k=1,\dots,d$ are the \textit{Tucker factors}, and $r_k$, $k=1,\dots,d$ are the \textit{Tucker ranks} of $\Aop$. Tucker decomposition provides an excellent approximation of $\Aop$, with an upper error bound guarantee, if computed with the higher\hyp order SVD (HOSVD) \cite{Lathauwer2000} or the Cross3D algorithm \cite{Oseledets2008, giannakopoulos20183d}. 

\section{Hybrid Volume-Surface Integral Equation}  
In this work, we propose a hybrid volume\hyp surface integral equation approach that combines pFFT \cite{guryev2019fast}, Tucker decomposition \cite{giannakopoulos2019memory}, TT-cross \cite{oseledets2010tt} and ACA \cite{Zhao2005} to optimize the assembly and the solution of the VSIE system.
\par
According to \cite{zhao2008domain}, the $2\times 2$ block system in \eqref{eq:nvsie} can be interpreted as the outcome of a domain decomposition method. Building on that, we can expand such system into a $3\times 3$ one, by separating the conductive surfaces into a ``near'' ($\rm n$) and a ``far'' ($\rm f$) domain, based on their distance from the sample. For a typical MRI simulation setup, the RF shield and the scanner's bore belong to the ``far'' domain, while the receive array is usually in the ``near'' domain (Fig. 1). The domain where transmit arrays are placed can vary, depending on their position relative to the body, and should be judiciously chosen for each application.

Applying the proposed domain decomposition, the system of equations in \eqref{eq:nvsie} becomes:
\begin{equation}
\begin{bmatrix}
Z^{\rm ff}_{\rm cc} & \!\left(Z^{\rm fn}_{\rm cc}\right)^{\rm T} & \!\left(Z^{\rm f}_{\rm cb}\right)^{\rm T} \\
Z^{\rm fn}_{\rm cc} & Z^{\rm nn}_{\rm cc}                & \!\left(Z^{\rm n}_{\rm cb}\right)^{\rm T} \\
Z^{\rm f}_{\rm cb}  & Z^{\rm n}_{\rm cb}                 & Z_{\rm bb}        
\end{bmatrix} \begin{bmatrix}
\matvec{j}^{\rm f}_{\rm c} \\
\matvec{j}^{\rm n}_{\rm c} \\
\matvec{j}_{\rm b} 
\end{bmatrix} = \begin{bmatrix}
\matvec{v}_{\rm f} \\
\matvec{v}_{\rm n} \\
0
\end{bmatrix}.
\label{eq:hvsie}
\end{equation}
Here, $\matvec{j}_{\rm c}^{\rm n}, \matvec{v}_{\rm n} \in \mathbb{C}^{m_n \times 1}$ and $\matvec{j}_{\rm c}^{\rm f}, \matvec{v}_{\rm f} \in \mathbb{C}^{m_f \times 1}$ are the unknown currents and voltage excitation vectors in the coils placed in the ``near'' and ``far'' domain, respectively. Note that $\matvec{v}_{\rm n}$ or $\matvec{v}_{\rm f}$ can be equal to zero, if no excitations exist in the corresponding domain. $Z^{\rm ff}_{\rm cc} \in \mathbb{C}^{m_f \times m_f}$ and $Z^{\rm nn}_{\rm cc} \in \mathbb{C}^{m_n \times m_n}$ are the Galerkin matrices that model the interactions between the $m_f$ ``far'' and the $m_n$ ``near'' surface discretization elements, respectively. $Z^{\rm fn}_{\rm cc} \in \mathbb{C}^{m_n \times m_f}$ is the off\hyp diagonal block of $Z_{\rm cc}$ and models the coupling interactions between conductors in the ``far'' and ``near'' domains. Since $Z^{\rm fn}_{\rm cc}$ has low\hyp rank properties, it can be compressed with the ACA method, as in \cite{Zhao2005}:
\begin{equation}
U V^* = \text{ACA} \!\left(Z^{\rm fn}_{\rm cc} \right).
\end{equation} 
Returning to \eqref{eq:hvsie}, $Z^{\rm n}_{\rm cb} \in \mathbb{C}^{m_n\times q \cdot n_v}$ and $Z^{\rm f}_{\rm cb} \in \mathbb{C}^{m_f \times q \cdot n_v}$ are the coupling matrices that model the interactions between the body and the conductive surfaces placed on the ``near'' and ``far'' domain, respectively. 
\par
The $2\times 2$ sub\hyp system in \eqref{eq:hvsie} that models the interactions between the body and the conductors placed in the ``near'' domain can be solved efficiently with the pFFT method \cite{guryev2019fast}:
\begin{equation}
Z^{\rm pFFT}_{\rm HOSVD} = \begin{bmatrix}
Z^{\rm nn}_{\rm cc} & \!\left(Z^{\rm n}_{\rm cb}\right)^{\rm T} \\
Z^{\rm n}_{\rm cb}  & Z_{\rm bb}
\end{bmatrix}.
\end{equation}
Precisely, the pFFT is used to project the mesh of the coil located in the ``near'' domain onto a uniform voxelized grid that encloses both the coil's triangular patches and the voxels that discretize the sample \cite{guryev2019fast}. The resulting $Z^{\rm pFFT}_{\rm HOSVD}$ matrix has a multilevel block\hyp Toeplitz structure and resembles the VIE integral equation matrix $Z_{\rm bb}$ of \eqref{eq:disc_vie}. The subscript HOSVD indicates that the Tucker decomposition method can be leveraged to compress the matrix and perform rapid multiplications in the case of fine voxel resolutions, as in \cite{giannakopoulos2019memory}.
\par
A previous study proposed performing a block\hyp wise assembly of $Z_{\rm cb}^{\rm f}$ and then compress each block with TT\hyp SVD \cite{giannakopoulos2021atensor} or HOSVD \cite{giannakopoulos2021compression}. In this work, instead of assembling the coupling matrix, we used a function that maps from indices to corresponding values in the matrix. Then, we calculated a low\hyp parametric approximation of the coupling matrix in $q$ 4D tensors, which required only a fraction of its entries, using a tensor decomposition and a cross approximation algorithm that iteratively calls the aforementioned function as a black\hyp box procedure. Precisely, we used the TT\hyp cross method \cite{oseledets2010tt} based on the adaptive density matrix renormalization group \cite{savostyanov2011fast}:
\begin{equation}
\begin{aligned}
\tilde{\Zop^{\chi}_{\rm cb}} &= \text{TT}\!\left( \Zop^{\chi}_{\rm cb} \right), \: \chi = 1,\dots,q, \\
\tilde{\Zop^{\chi}_{\rm cb}}_{,\: i_1 i_2 \cdots i_4} &= \sum\limits_{j_1,j_2,j_3}^{r_1,r_2,r_3} G^1_{i_1 j_1} \Gop^2_{j_1 i_2 j_2} \Gop^{3}_{j_2 i_3 j_3} G^4_{j_3 i_4},
\end{aligned}
\label{eq:ass_TT} 
\end{equation} 
where $\Zop^{\chi}_{\rm cb}$ is the reshaped $\chi$-th 4D tensor component of $Z^{\rm f}_{\rm cb}$, and $\tilde{\Zop^{\chi}_{\rm cb}}$ is the TT compressed $\Zop^{\chi}_{\rm cb}$ that is computed using \eqref{eq:TT}. Note that we never fully assemble $\tilde{\Zop^{\chi}_{\rm cb}}$, but we rather express it using the cores of its TT: $G^1,\Gop^2,\Gop^3$, and $G^4$. 
\par
By combining all the aforementioned methods, we can transform \eqref{eq:hvsie} into a hybrid\hyp VSIE system with low memory footprint that can be solved rapidly: 
\begin{equation}
\scriptstyle
\begin{bmatrix}
Z^{\rm ff}_{\rm cc}                                                       & \bar{V} U^{\rm T} & \left[\tilde{\Zop^{1}_{\rm cb}},\vdots,\tilde{\Zop^{q}_{\rm cb}} \right]^{\rm T}          \\
U V^*                                                                     &             &                                                                                     \\
\left[\tilde{\Zop^{1}_{\rm cb}},\vdots,\tilde{\Zop^{q}_{\rm cb}} \right]  & \multicolumn{2}{c}{\raisebox{\ht\strutbox-.4\height}[0pt][0pt]{$Z^{\rm pFFT}_{\rm HOSVD}$}}
\end{bmatrix} \begin{bmatrix}
\matvec{j}^{\rm f}_{\rm c} \\
\matvec{j}^{\rm n}_{\rm c} \\
\matvec{j}_{\rm b} 
\end{bmatrix} = \begin{bmatrix}
\matvec{v}_{\rm f} \\
\matvec{v}_{\rm n} \\
0
\end{bmatrix}.
\label{eq:hvsie_disc}
\end{equation}

\subsection{Solving the hybrid-VSIE} 
Since it is not practical to directly invert the hybrid\hyp VSIE matrix in \eqref{eq:hvsie_disc}, even for coarse resolutions, an iterative solver, like the generalized minimal residual method (GMRES) \cite{saad1986gmres}, should be used. Multiplications involving $Z^{\text{pFFT}}_{\text{HOSVD}}$ can be carried out efficiently using the methods presented in \cite{giannakopoulos2019memory}. Matrix\hyp vector products involving $Z^{\rm ff}_{\rm cc}$, $U V^*$, and $\bar{V} U^{\rm T}$ can be performed with standard matrix multiplications. For the multiplications involving the TT\hyp compressed $Z^{\rm f}_{\rm cb}$ we propose the following approach.
\par 
Let us consider the product $[\tilde{\Zop^{1}_{\rm cb}},\smash{\vdots},\tilde{\Zop^{q}_{\rm cb}}] \matvec{j}_{\rm c}^{\rm f}$. We can re-write it as: 
\begin{equation}
\left[\tilde{\Zop^{1}_{\rm cb}},\vdots,\tilde{\Zop^{q}_{\rm cb}}\right] \matvec{j}_{\rm c}^{\rm f} = \left[\tilde{\Zop^{1}_{\rm cb}} \matvec{j}_{\rm c}^{\rm f},\vdots,\tilde{\Zop^{q}_{\rm cb}} \matvec{j}_{\rm c}^{\rm f}\right]. 
\end{equation}
We can decompose $\tilde{\Zop^{\chi}_{\rm cb}}$ using \eqref{eq:ass_TT} and then perform the product $\tilde{\Zop^{\chi}_{\rm cb}} \matvec{j}_{\rm c}^{\rm f}$ in a multiple step process, for each $\chi \in {1,\dots,q}$. First we can reshape $\Gop^2 \in \mathbb{C}^{r_1 \times n_2 \times r_2} \rightarrow G^2 \in \mathbb{C}^{(r_1 \cdot n_2) \times r_2}$ and $\Gop^3 \in \mathbb{C}^{r_2 \times n_3 \times r_3} \rightarrow G^3 \in \mathbb{C}^{(r_2 \cdot n_3) \times r_3}$, where $\rightarrow$ represents the reshape operation. Then, we can perform the following steps.
\begin{equation}
\begin{aligned}
\matvec{y} &= G^4 \matvec{j}_{\rm c}^{\rm f}, \:\: \matvec{y} \in \mathbb{C}^{r_3 \times 1},                                                                                            \\
\matvec{y} &= G^3 \matvec{y}, \:\: \matvec{y} \in \mathbb{C}^{r_2 \cdot n_3 \times 1}, \:\:\:\:\: \matvec{y} \rightarrow Y \in \mathbb{C}^{r_2 \times n_3},             \\ 
Y &= G^2 Y,                   \:\: Y \in \mathbb{C}^{r_1 \cdot n_2 \times n_3},        \:\: Y \rightarrow Y \in \mathbb{C}^{r_1 \times n_2 \cdot n_3},                  \\
Y &= G^1 Y,                   \:\: Y \in \mathbb{C}^{n_1 \times n_2 \cdot n_3},        \:\: Y \rightarrow \matvec{y} \in \mathbb{C}^{n_1 \cdot n_2 \cdot n_3 \times 1}.
\end{aligned}
\label{eq:mvp}
\end{equation}
Note that we implemented first the product $G^4 \matvec{x}$ to eliminate the fourth and largest dimension of $\Zop^{\chi}_{\rm cb}$, in order to reduce the memory footprint and the operation cost of the remaining multiplications.
\par
To compute the product $[\tilde{\Zop^{1}_{\rm cb}},\smash{\vdots},\tilde{\Zop^{q}_{\rm cb}}]^{\rm T} \matvec{j}_{\rm b}$, we applied the transpose operator:
\begin{equation}
\matvec{j}_{\rm b}^{\rm T} \left[\tilde{\Zop^{1}_{\rm cb}},\vdots,\tilde{\Zop^{q}_{\rm cb}}\right] = \sum\limits_{\chi = 1}^q \!\left(\!\left(\matvec{j}_{\rm b}^{\chi}\right)^{\rm T} \tilde{\Zop^{\chi}_{\rm cb}} \right)^{\rm T},
\label{eq:tmvp}
\end{equation}
where we separated $\matvec{j}_{\rm b}$ into its vector components $\matvec{j}_{\rm b} = [\matvec{j}_{\rm b}^1,\smash{\vdots},\matvec{j}_{\rm b}^q]$, with each $\matvec{j}_{\rm b}^{\chi} \in \mathbb{C}^{n_1\cdot n_2\cdot n_3 \times 1}$. The multiplications in \eqref{eq:tmvp} can be broken down into concatenated steps as in \eqref{eq:mvp}:
\begin{equation}
\begin{aligned}
\matvec{j}_{\rm b}^{\chi} & \rightarrow Y \in \mathbb{C}^{n_3 \cdot n_2 \times n_1}, \\
Y &= Y G^1,                   \:\: Y \in \mathbb{C}^{n_3 \cdot n_2 \times r_1}, \:\: Y \rightarrow Y \in \mathbb{C}^{n_3 \times r_1 \cdot n_2},  \\
Y &= Y G^2,                   \:\: Y \in \mathbb{C}^{n_3 \times r_2}, \:\:\:\:\:\:\: Y \rightarrow \matvec{y} \in \mathbb{C}^{1 \times n_3 \cdot r_2},     \\
\matvec{y} &= \matvec{y} G^3, \:\: \matvec{y} \in \mathbb{C}^{1 \times r_3},                                                                     \\ 
\matvec{y} &= \matvec{y} G^4, \:\: \matvec{y} \in \mathbb{C}^{1 \times m_f}.       
\end{aligned}
\end{equation}

\section{Methods}
\subsection{Realistic Head Models}
We evaluated the performance of our proposed hybrid-VSIE using the realistic Billie and Duke head models from the virtual family population \cite{VirtualFamily}. Both models were discretized over a regular voxelized grid. Billie was discretized with $5$, $2$, and $1$ mm voxel isotropic resolutions (VIR), corresponding to a simulation domain of $34 \times 38 \times 45$, $84 \times 96 \times 116$, and $168 \times 188 \times 222$ voxels, respectively. Duke was discretized with a $5$ mm VIR ($38 \times 47 \times 46$ voxels).  

\subsection{Simulation Setups}
\subsubsection{Triangular head coil}
We modeled the $7$ tesla transceiver triangular head array introduced in \cite{giannakopoulos2020magnetic} with a surrounding cylindrical perfectly electric conducting shield, loaded with Billie head model. The shield had a $20$ cm radius, was $35$ cm long, and was placed $1.7$ cm away in the $\hat{z}$ direction from the bottom end ring of the coil array as shown in Fig. 1. Details on the coil's geometry, tuning, and decoupling can be found in \cite{giannakopoulos2020magnetic}. For the case when Billie was discretized with $5$ and $2$ mm VIR, the array and the RF shield were discretized with $2140$ and $2662$ triangular elements, respectively. When Billie was discretized with $1$ mm VIR, the RF shield's discretization remained the same, whereas the array's mesh was refined to $6054$ triangular elements. The discretization of the conductors was adjusted based on the body resolution because for pFFT to be applicable, each pair of triangular elements must be contained by the voxel expansion domain of pFFT, which was set to $5\times 5\times 5$ voxels.

\renewcommand{\thefigure}{1}
\begin{figure}[ht!]
\begin{center}
\includegraphics[width=0.48\textwidth, trim={0 0 0 0}]{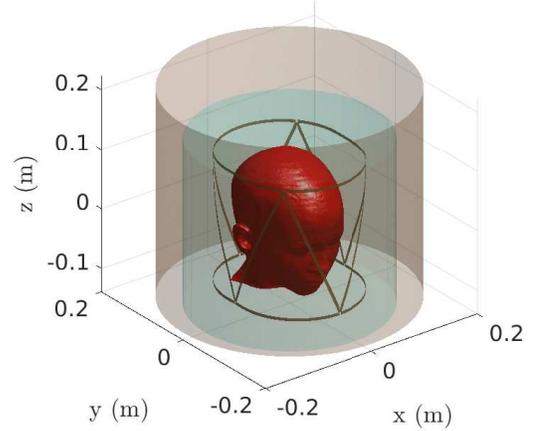}
\caption{Shielded 8\hyp element triangular coil array loaded with the Billie head model. The cyan volume defines the domain ``near'' the sample, which contains the coil, whereas the shield belongs to the ``far'' domain, outside the cyan region}.
\label{fig:n1}
\end{center}
\end{figure} 

\subsubsection{Single loop}
To qualitatively replicate previous results obtained with a commercial EM simulation software \cite{zhang2021effect}, we modeled a single surface loop placed behind the Duke head model (Fig. 2) at $3$, $7$, $9.4$, and $10.5$ tesla. The radius of the loop was $4.15$ cm, the conductor's width was $0.3$ cm and was discretized with $1958$ elements. The loop was segmented with seven capacitors for tuning and one capacitor connected in parallel to the feeding port for matching. The loop and head were surrounded by a $50$ cm long RF shield, while the radius was varied from $17.25$ to $32$ cm ($16$ positions, $1$ cm steps), for which the number of discretization elements ranged from $3264$ to $5824$. 

\renewcommand{\thefigure}{2}
\begin{figure}[ht!]
\begin{center}
\includegraphics[width=0.24\textwidth, trim={0 0 0 0}]{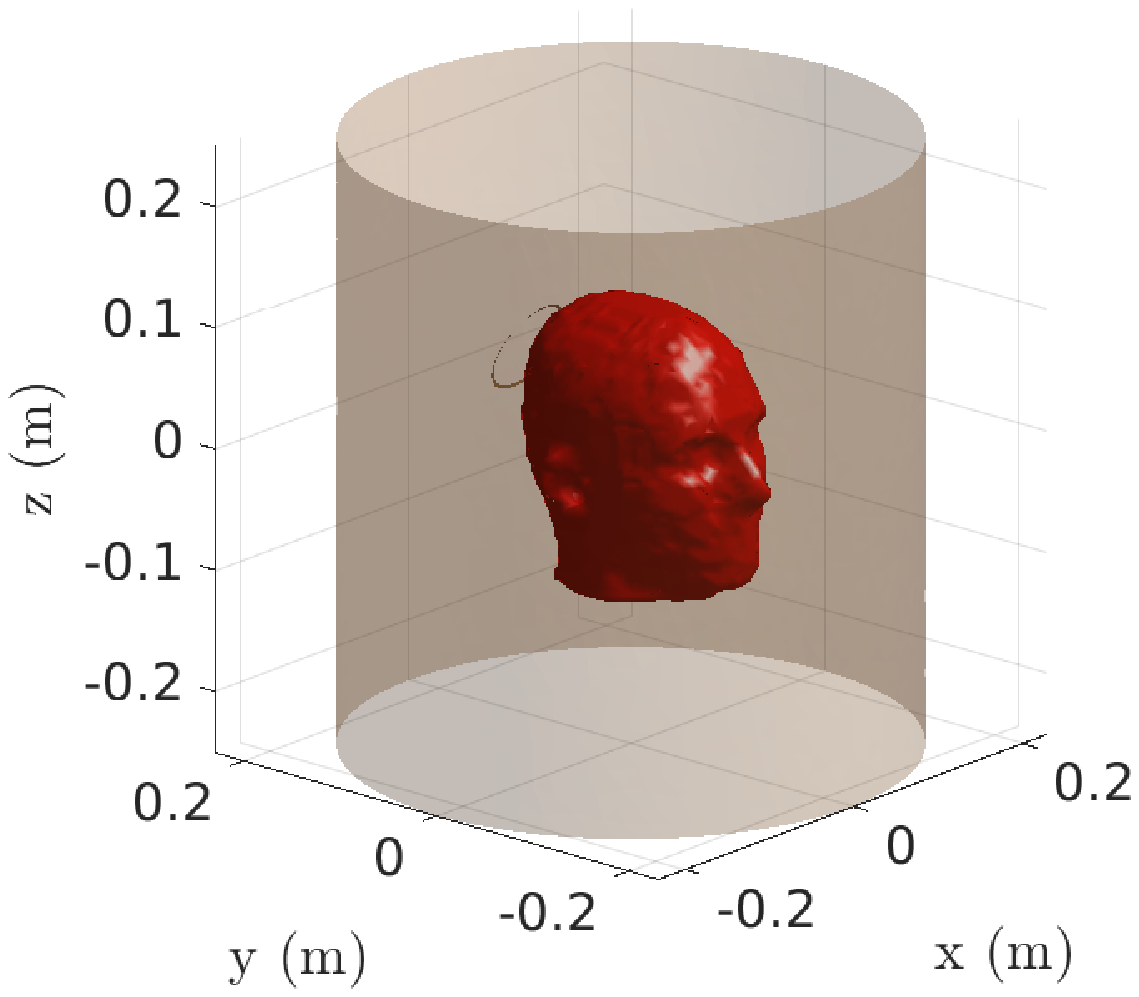}
\includegraphics[width=0.24\textwidth, trim={0 0 0 0}]{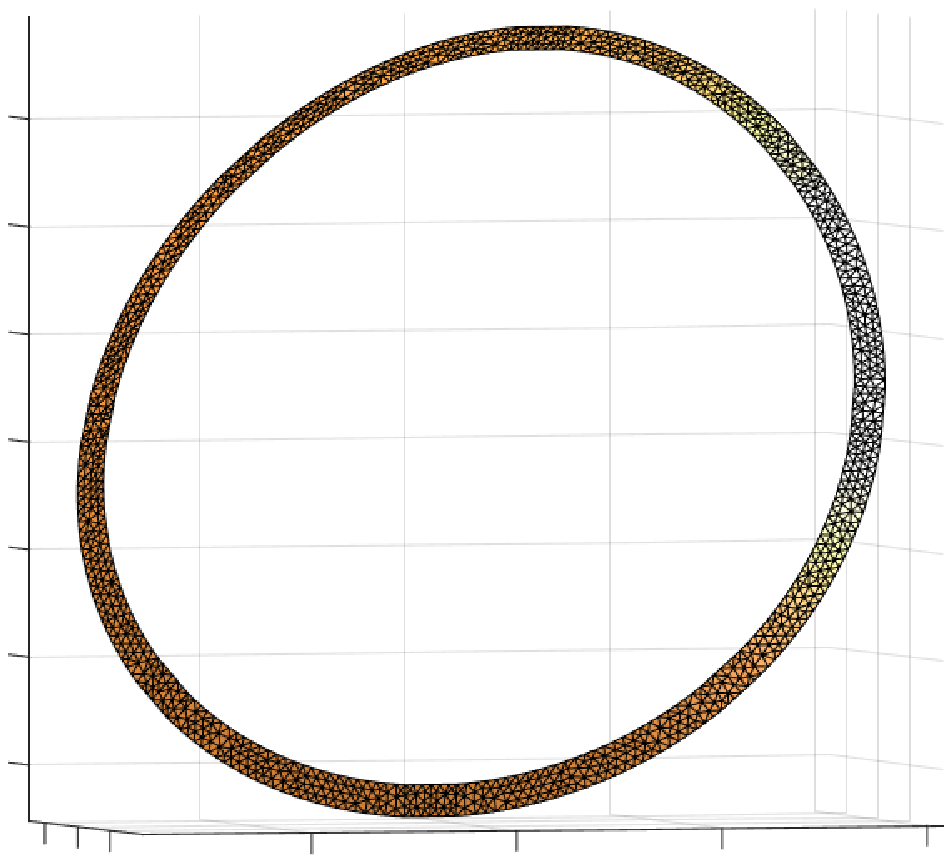}
\caption{(left) RF shield encircling a loop loaded with the Duke head model. (right) Details of the loop's geometry. The figure shows a representative RF shield radius.}
\label{fig:n2}
\end{center}
\end{figure}

\subsubsection{Overlapping transceiver array}
We modeled an eight\hyp channel transceiver array with overlapping coil elements (Fig. 3) at $9.4$ tesla. The array consisted of two sets of four identical rectangular loops of $11.88$ cm width and $8$ cm length, arranged on two cylindrical surfaces of length $8$ cm, and radii $12$ and $12.5$ cm. The four loops of each set were interleaved by $45$ degrees. The neighboring coils overlapped by $3$ cm to ensure geometric decoupling. The coil conductors were $0.5$ cm wide and were discretized with $2792$ triangular elements. Each coil was segmented with eight variable capacitors for tuning and one variable capacitor connected in parallel at the feeding port for matching. The array was surrounded by the same RF shields as in the single-loop setup and also loaded with the Duke head model. For this setup, we modeled also the scanner's bore as an open cylindrical surface of $277.75$ cm length and $32.05$ cm radius, which was discretized with $5596$ triangular patches.

\renewcommand{\thefigure}{3}
\begin{figure}[ht!]
\begin{center}
\includegraphics[width=0.24\textwidth, trim={0 0 0 0}]{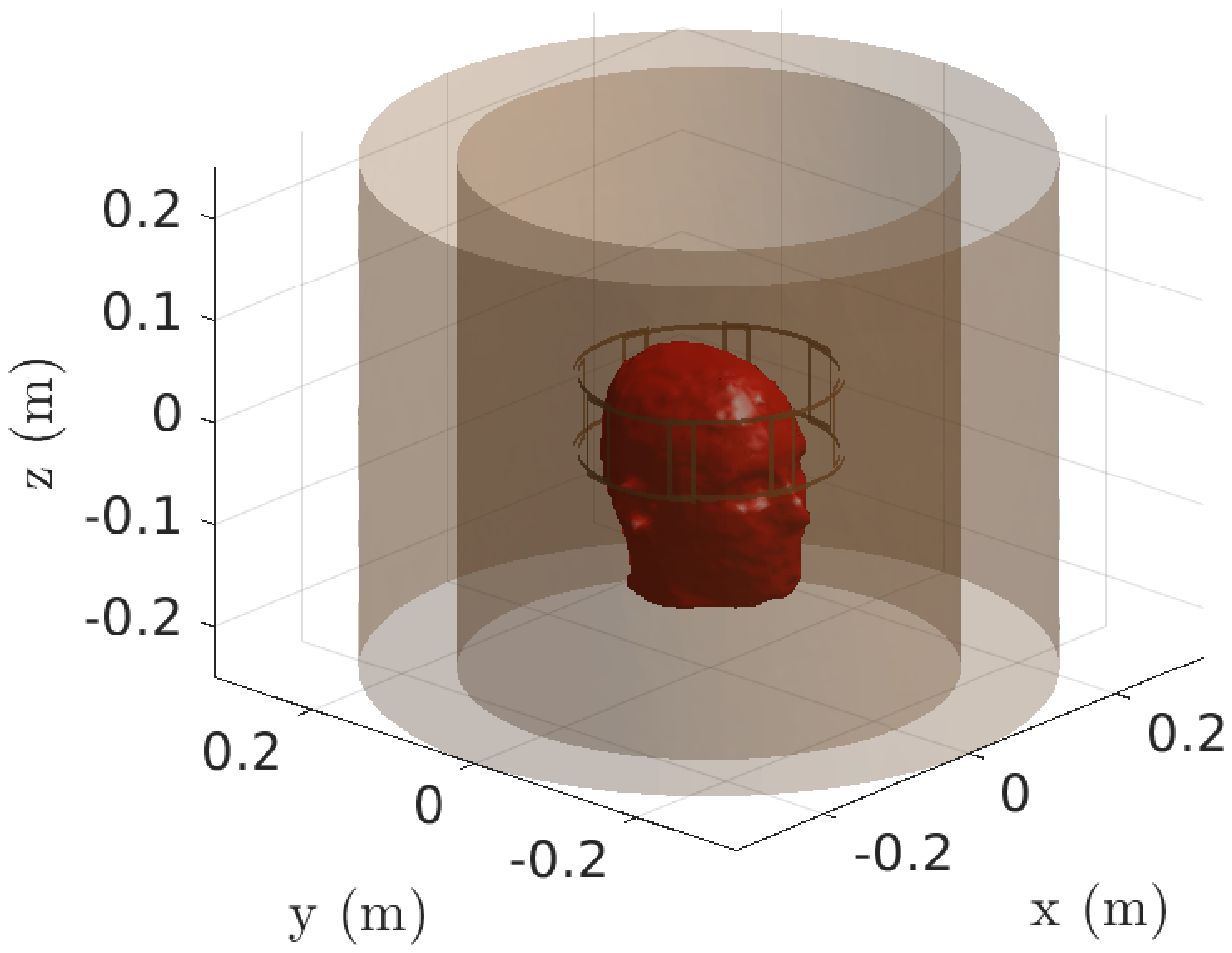}
\includegraphics[width=0.24\textwidth, trim={0 0.5cm 0 0}]{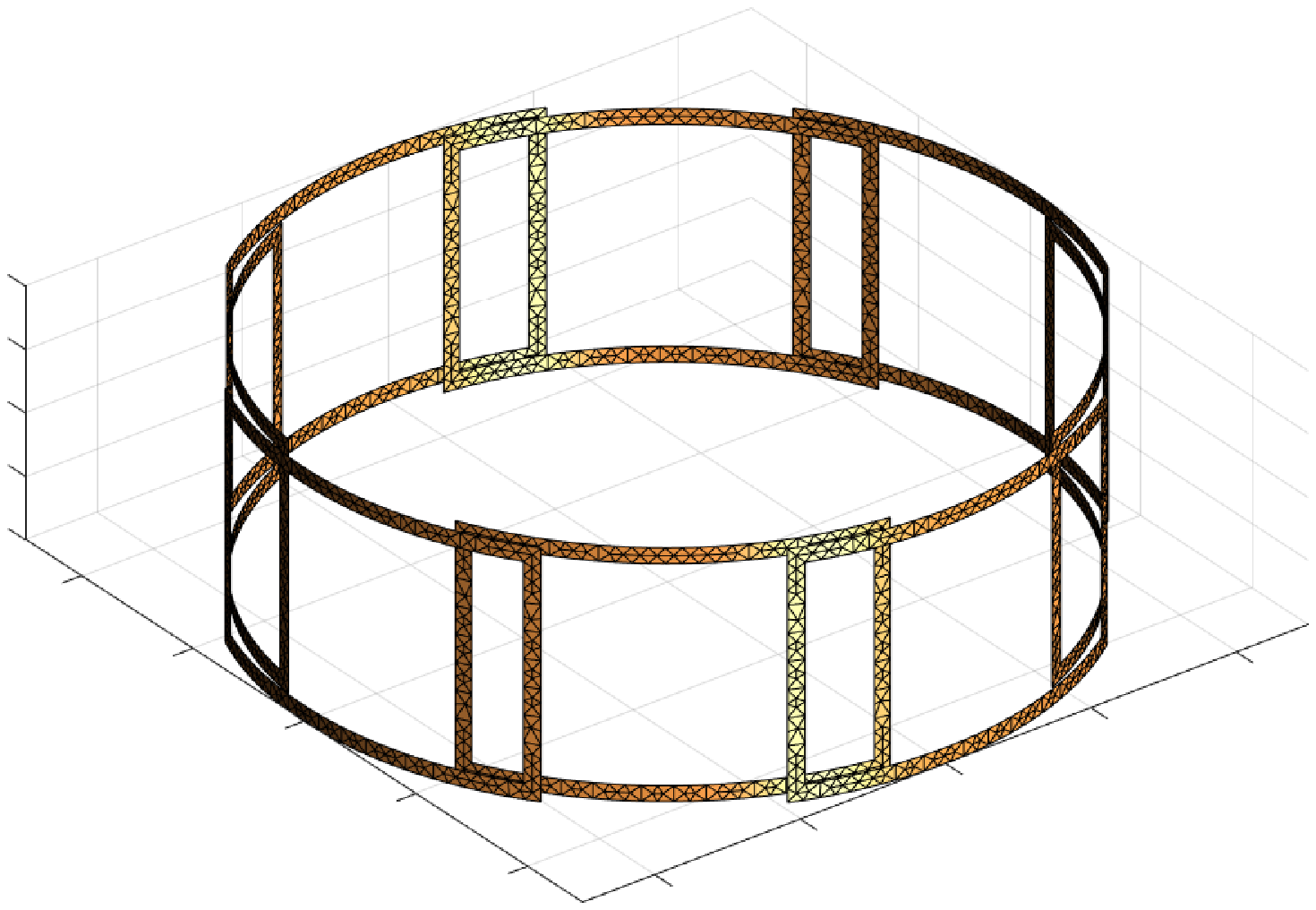}
\caption{(left) Transceiver array loaded with the Duke head, surrounded by an RF shield and the scanner's bore. Note that the bore's length is cropped in the figure for enhanced visualization and a representative radius is displayed for the RF shield. (right) Details of the array's geometry.}
\label{fig:n3}
\end{center}
\end{figure}

\subsection{Numerical Experiments}
All simulations were performed on a server running Ubuntu 18.04.2 LTS operating system, with an Intel(R) Xeon(R) Gold 6248R CPU at 2.70GHz, 112 cores, 2 threads per core, and an NVIDIA A100 PCIe GPU with 40GB of memory. All matrix\hyp vector products were implemented in MATLAB 9.10 (The MathWorks Inc., Natick, MA).

\subsubsection{Comparison with Non-Hybrid Methods}
We compared our proposed hybrid\hyp VSIE (H) method with three non\hyp hybrid\hyp VSIE approaches that solved the system of equations \eqref{eq:nvsie}, for the setup in Fig. 1. For the first non\hyp hybrid method (I), the coupling matrix between the coil array, the shield, and the head model was assembled in its full form. For the second method (II), the coupling matrix was assembled in a compressed form with the ACA method. Note that we used methods I and II for the comparison because they are implemented in MARIE. For the third method (III), pFFT was used to project both the coil and the shield to an extended voxelized domain as in \cite{guryev2019fast}. For the hybrid\hyp VSIE method, the coil, tightly-fitting the head model, was placed in the ``near'' domain, while the shield was placed in the ``far'' domain. The polarization body currents were modeled with PWC and PWL basis functions for comparison. 
\par
For all simulations, the system was solved with the GMRES, executed either on GPU (G) or CPU (C) depending the memory requirements. The tolerance was set to $10^{-5}$. GMRES' restarts were set to $50$ for all cases, except for the $1$ mm VIR\hyp PWL basis functions case, where they were reduced to $10$ to avoid GPU memory overflows. The tolerances for TT\hyp cross and ACA in the hybrid\hyp VSIE method were $10^{-3}$. The ACA tolerance in method II was set to $1$\hyp $2$ orders of magnitude lower than the GMRES tolerance. The HOSVD tolerance was $10^{-5}$ for all methods. For methods I,II, and III, we used $8$ volume and $5$ surface quadrature integration points to approximate the coupling matrix integrals. In the hybrid\hyp VSIE, we used $8$ and $1$ volume, and $5$ and $1$ surface integration points for the computations of $Z_{\rm bc}^{\rm n}$ and $Z_{\rm bc}^{\rm f}$, respectively.

\subsubsection{Effect of the Shield}
We used the hybrid\hyp VSIE method to study the effect of the RF shield's position on the SNR distribution. For the loop in Fig. 2, we repeated $16$ simulations with the hybrid\hyp VSIE (one for each shield's radius) for four main magnetic field strengths and both PWC and PWL basis functions at $5$ mm VIR (128 simulations in total). For the array in Fig. 3, we used the hybrid\hyp VSIE method to perform simulations at $9.4$ tesla for all $16$ shield's radii, both with and without the scanner's bore. To complement geometric decoupling between first-order neighbors, the capacitor values were adjusted to further reduce the relative scattering (S) parameters. We calculated the SNR \cite{lattanzi2010performance} and averaged its value over all voxels of the head. Both the loop and the multiple channel array were re\hyp tuned and matched for all shield's positions and all field strengths. Coil elements were placed in the ``near'' domain, while the RF shield and the scanner's bore were in the ``far'' domain. The polarization body currents were modeled with PWC and PWL basis functions for the single\hyp loop case, and with PWC only for the array case. We used the same tolerances and number of quadrature points as in the first experiment.

\section{Numerical Results}
\subsection{Comparison with Non-Hybrid Methods}  
TABLE I compares the convergence time and iterations ($\ceil*{\rm hh:mm:ss}$ - $\ceil*{\rm iter}$), averaged over the eight solutions of the system (one solution for each channel of the triangular coil), for the hybrid\hyp VSIE against methods I, II, and III. The bold font indicates the faster solver in each case. Methods I, II, and III were not applicable in all cases, either due to memory limitations or because the resulting VSIE matrix from \eqref{eq:nvsie} was ill\hyp conditioned. 

\begin{table}[ht!]
\caption{Average GMRES Convergence Properties} \label{tb:times} \centering
{\def\arraystretch{1.3}\tabcolsep=6.5pt
\begin{tabular}{ c c c c c }
\hline
\multirow{2}{*}{VIR}    & \multirow{2}{*}{VSIE}  & PWC                         & PWL                          \\
                        &                        & hh:mm:ss - iter - PU        & hh:mm:ss - iter - PU         \\
\hline
\hline
\multirow{4}{*}{$5$ mm} & H                      & \textbf{00:00:03} - 136 - G & \textbf{00:00:19} - 165 - G  \\
						& I                      & 00:00:06 - 136 - G          & 00:04:25 - 165 - C           \\
						& II                     & 00:00:04 - 136 - G          & 00:00:22 - 165 - G           \\
						& III                    & 00:00:04 - 170 - G          & 00:00:24 - 199 - G           \\
\hline						
\multirow{3}{*}{$2$ mm} & H                      & \textbf{00:00:16} - 177 - G & \textbf{00:01:37} - 193 - G  \\
						& I                      & 00:18:45 - 177 - C          & N/A                          \\
						& II                     & 00:06:17 - 177 - C          & N/A                          \\
\hline						
                $1$ mm  & H                      & \textbf{00:02:57} - 192 - G & \textbf{01:49:12} - 327 - G  \\
\hline \\
\end{tabular}
}
\end{table}

TABLE II shows the $\rm L_2$ norm relative differences between the solutions of methods H, II, and III with respect to method I, which was assumed as the gold-standard because it uses the fully assembled system. The comparison was possible only for the $5$ mm VIR and both PWC and PWL basis functions, and $2$ mm VIR using only the PWC basis. In the remaining cases, method I could not be implemented on our server due to memory overflow.

\begin{table}[ht!]
\caption{Relative differences with respect to Method I} \label{tb:errors} \centering
{\def\arraystretch{1.3}\tabcolsep=6.5pt
\begin{tabular}{ c c c c }
\hline
Method     & PWC $5$ mm & PWL $5$ mm & PWC $2$ mm  \\
\hline
\hline
H          & $0.08\%$   & $2.76\%$   & $0.52\%$    \\
II         & $0.36\%$   & $0.50\%$   & $0.95\%$    \\
III        & $0.42\%$   & $0.38\%$   & N/A         \\
\hline						
\end{tabular}
}
\end{table}

A qualitatively comparison between PWC and PWL basis functions is shown in Fig. 4 for $1$ mm VIR, for the circularly polarized (CP) mode of the array, in which the eight transmit fields of the coil elements were combined to cancel the phase at the center of the head.

\renewcommand{\thefigure}{4}
\begin{figure}[ht!]
\begin{center}
\includegraphics[width=0.48\textwidth, trim={0 0 0 0}]{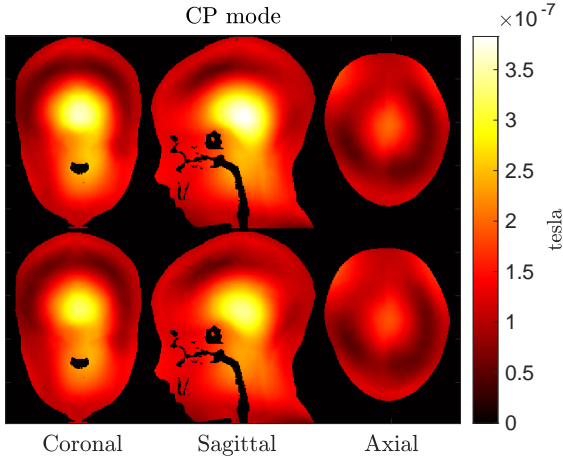}
\caption{Magnitude of transmit field of the triangular coil array in CP mode combination. Results are shown for $1$ mm VIR, for PWC (top) and PWL (bottom) basis functions, and three representative orthogonal slices of the head model. The values are masked outside Billie for an enhanced visualization.}
\label{fig:n4}
\end{center}
\end{figure} 

\subsection{Effect of the Shield}
\subsubsection{Single Loop}
The left axis in Fig. 5 shows the normalized average value of the SNR inside the head as a function of the RF shield's diameter. The right axis shows the corresponding compression factor of $Z_{\rm bc}^{\rm f}$, computed as the number of elements of the full matrix divided by the number of elements of the TT\hyp compressed one. 

\renewcommand{\thefigure}{5}
\begin{figure*}[ht!]
\begin{center}
\includegraphics[width=1.0\textwidth, trim={0 0 0 0}]{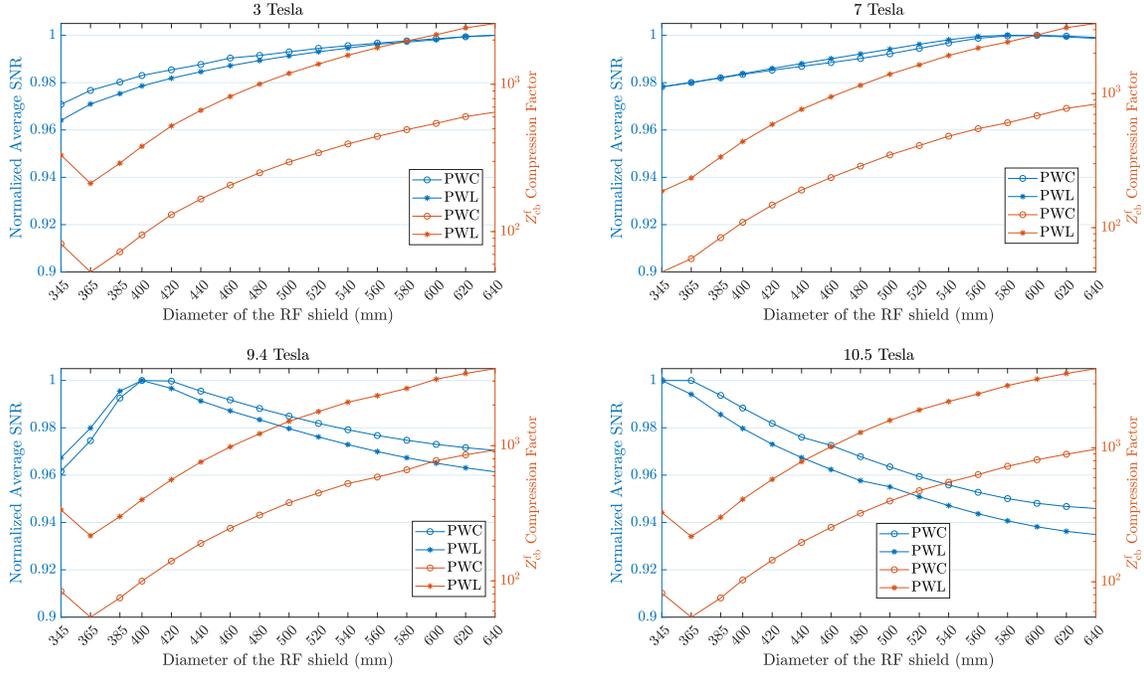}
\caption{(left axis, blue) Normalized average SNR as a function of the radius of the RF shield for different main magnetic field strengths and basis functions (PWC and PWL). (right axis, red) Compression factor of $Z_{\rm bc}^{\rm f}$. for each case. Note that the loop was re\hyp tuned for each simulation.}
\label{fig:n5}
\end{center}
\end{figure*}

\subsubsection{8-element Array}
The worst S parameter between first-order neighbors was $-17.3$ dB for $620$ mm shield diameter (w/o the bore), whereas coupling was slightly higher between next-nearest and opposite neighbors, with the worst case being $-12.7$ dB between two opposite coils for $640$ mm shield diameter (w/o the bore). Fig. 6 shows the normalized average SNR (left axis) and the compression factor of $Z_{\rm bc}^{\rm f}$ (right axis), as a function of the shield's diameter.

\renewcommand{\thefigure}{6}
\begin{figure}[ht!]
\begin{center}
\includegraphics[width=0.48\textwidth, trim={0 0 0 0}]{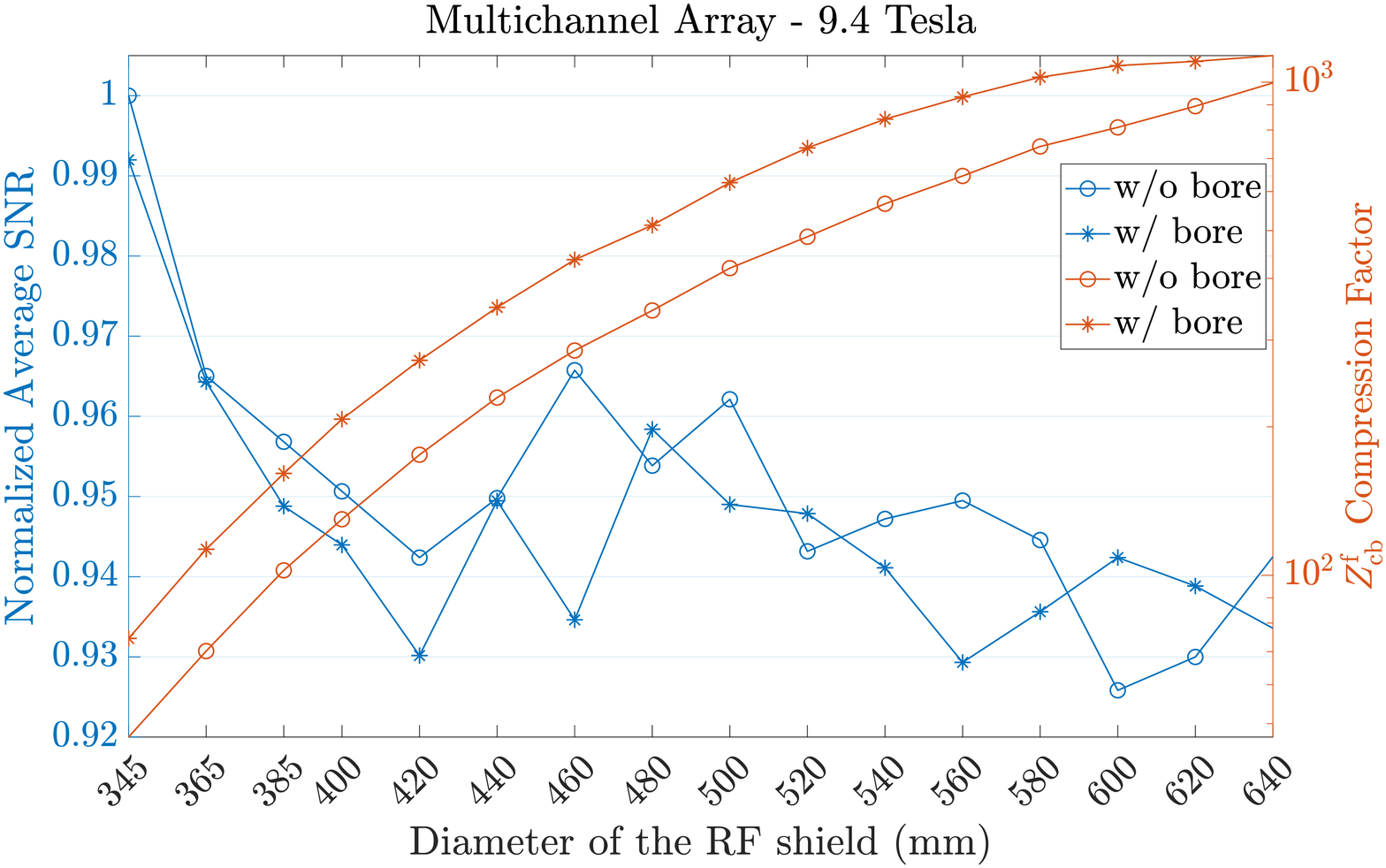}
\caption{Normalized average SNR of the head voxels in respect to the radius of the RF shield. The array was re\hyp tuned and decoupled for all cases.}
\label{fig:n6}
\end{center}
\end{figure}

\section{Discussion}
The aim of this work was to introduce a hybrid\hyp VSIE technique that exploits the low\hyp rank properties of the memory\hyp intensive parts of the VSIE to enable rapid EM simulations for realistic MRI scenarios. The separation of the conductive surfaces into a ``near'' and a ``far'' domain yielded exceptional numerical performances for the hybrid\hyp VSIE case. In particular, the cross\hyp TT considerably reduced the memory footprint of the $Z^{\rm f}_{\rm cb}$ matrix. This is useful when the ``far'' domain is represented by an RF shield, since the compressed matrix can be stored and recycled for multiple EM simulations involving different RF coils (for example, the same $Z^{\rm f}_{\rm cb}$ could be used for the results in V.B.1 and V.B.2), or inside coil design optimization pipelines \cite{seralles2018parametric}. 
\par
While one could assemble $Z^{\rm f}_{\rm cb}$ using SVD or ACA instead of TT for compression, the memory requirements would be larger \cite{giannakopoulos2021compression}. TT\hyp cross was preferred over TT\hyp SVD, because it does not require the full or a block\hyp wise assembly of $Z^{\rm f}_{\rm cb}$ prior to the decomposition - a task that can be lengthy for fine VIR or PWL basis functions \cite{giannakopoulos2021atensor}. We chose to use ACA in method II because such approach is implemented in MARIE. Alternatively, one could use TT in method II to assemble the whole coupling matrix $Z_{\rm cb}$ as in \cite{giannakopoulos2021atensor}. However, the assembly time of the compressed matrix $Z_{\rm cb}$, and the time to compute the relevant matrix\hyp vector products \eqref{eq:mvp}, \eqref{eq:tmvp} could increase, because the local interactions between the conductive surfaces and the body could lead to higher matrix ranks, therefore the hybrid-VSIE was deemed a superior choice. Further investigation is needed when the conductive surfaces are neither near nor far from the body, for example, when simulating body\hyp transmit coils. In these cases, as in setups where all conductive surfaces are near or far, the hybrid\hyp VSIE would not be advantageous and one could use either pFFT or TT\hyp cross on \eqref{eq:nvsie} instead. The criteria to choose the optimal method would depend on the distance of the coil from the body, and the discretization chosen for the body and the coil. In future work, one could also explore other approaches to achieve further acceleration, for example, the use of the Calder\'{o}n preconditioner \cite{andriulli2008multiplicative} to reduce the number of iterations of GMRES.
\par
The hybrid\hyp VSIE was the only practical method in the case of fine VIR. Method III required a larger number of iterations for the convergence of GMRES even in the case of $5$ mm VIR, because each triangular pair of the RF shield was not small enough to be projected accurately onto the $5\times 5\times 5$ voxel expansion domain of pFFT \cite{phillips1997precorrected, guryev2019fast}, resulting in a ill\hyp conditioned system. One way to tackle this would be to enlarge the expansion domain of pFFT, at the cost of additional assembly and solution times. At $1$ mm VIR, the hybrid\hyp VSIE became ill\hyp conditioned like method III for the chosen pFFT expansion domain, thus GMRES, required a larger number of iterations to converge. To address this, instead of enlarging pFFT's expansion domain, we refined the triangular mesh of the array, leading to an $\sim\! 8$ times larger $Z_{\rm cc}^{\rm nn}$ matrix. Note that in method III, the same refinement would increase the size of $Z_{\rm cc}$ by $\sim\! 1400$ times. Methods I and II required additional memory for the coupling matrix assembly and multiplications, so they became slow or impractical for fine VIR. In the case of $2$ mm VIR resolution and PWL basis functions, one could assemble the coupling matrix $Z_{\rm cb}$ in methods I and II using a point-matching integration (thus, neglecting the contribution of the linear terms), instead of the Petrov\hyp Galerkin approximation, reducing its memory requirements by $4$, at the cost of lower accuracy in the solution.   
\par
The hybrid\hyp VSIE allowed us to use higher tolerances and fewer quadrature integration points for the Galerkin matrices that involved remote interactions of the body with conductors in the ``far'' domain. This was possible because such interactions are less pronounced than the ones in the ``near'' domain. As a result, the hybrid\hyp VSIE method remained accurate ($2.76\%$ relative difference of the solution with respect to method I) for PWL basis functions, even though the contribution of the linear terms in $Z_{\rm bc}^{\rm f}$ was neglected. By using more quadrature points to include linear term contributions, the difference becomes smaller than $0.1\%$. 
\par
TABLE I shows that the PWL approximation required additional time to converge to the desired tolerance. Nevertheless, as suggested by Fig. 3, PWL are needed for accurate modeling of the magnetic field density \cite{georgakis2020fast}. Such level of accuracy may not be needed for qualitative coil assessment, but is necessary for other quantitative applications \cite{serralles2019noninvasive}.
\par
The marked improvement of the convergence time of hybrid\hyp VSIE allowed us to perform optimization of the RF shield's position with respect to SNR. In section V.B.1, we verified the accuracy of hybrid\hyp VSIE by repeating the same series of numerical simulations reported in \cite{zhang2021effect}. The loop's and shield's positions and geometries were exactly the same in both series of simulations, while the head phantom was placed $1$ cm higher in the hybrid\hyp VSIE simulation. For the $3$T and $7$T cases, the results matched, with minor inconsistencies. For $9.4$T and $10.5$T, the behavior of the average SNR as a function of the shield diameter was similar, except for the location of the peak values, which occurred for a $30$ mm smaller shield radius for the hybrid\hyp VSIE. Such minor discrepancies are expected, since in \cite{zhang2021effect} the EM field was computed with a different approach based on the finite integration technique.
\par
In section V.B.2 we used the hybrid\hyp VSIE to study the performance of a multi-channel array with respect to the RF shield's size at $9.4$ tesla. In particular, we observed approximately $7\%$ difference between the highest ($345$ mm shield diameter) and lowest ($600$ mm shield diameter) average SNR value. The inclusion of the scanner's bore in the simulations had a negligible effect on the average SNR ($\leq 3\%$). Note that, as in the single loop case, the SNR value did not change monotonically with the shield diameter at $9.4$ tesla. Note that in the case of the array, the behavior of the average SNR as a function of shield diameter was not smooth. We believe this is mainly due to two reasons: 1) the RF interference patterns generated by the array elements change based on the distance from the shield to the coils, as shown in \cite{zhang2021effect} 2) the tuning and coupling of each coil element were not identical for every case. Specifically, after re\hyp tuning for each shield position, the S parameters of the array elements varied by up to a few dB, which affected the resulting fields. This effect was instead negligible for the single loop, since the S parameter was $\ll -50$ dB for all shield diameters.
\par
The hybrid\hyp VSIE is a domain\hyp decomposition method and therefore carries the limitations of the methods used to model each domain. Specifically, pFFT requires that each triangular pair of the conductors in the ``near'' domain fits in the pFFT voxel expansion domain in order to be performed. TT\hyp cross algorithm might overestimate the TT\hyp ranks, which might lead to unnecessary additional memory allocation. Note that this is not a practical limitation for the simulation's performance, since the coupling matrix could still be compressed by a large factor with TT\hyp cross. However, if one is interested in assembling a large dataset consisting of multiple $Z^{\rm f}_{\rm cb}$ matrices for RF shield optimization pipelines, the overestimated TT\hyp ranks might lead to storage concerns. Finally, since the equivalent shield and coil currents and the polarization body currents are solved in a coupled fashion in our approach, the choice of the tolerance value for the different techniques used in hybrid\hyp VSIE is important to avoid false convergence of the iterative solver. In fact, since the coil currents are normally larger than the body and shield currents, by using a single tolerance, the accuracy of the latter could be lower than the accuracy of the coil currents. While this is expected to have a negligible effect on coil simulations, one could ensure the same accuracy by solving first for the polarization currents, and then computing the equivalent coil currents, as a post-processing step. Such approach would allow further flexibility in the choice of the iterative's solver tolerance.

\section{Conclusion}
We presented a new hybrid\hyp VSIE method suitable for rapid and comprehensive EM simulations in MRI. Our proposed method combines TT, pFFT, Tucker decomposition, and the ACA to heavily compress the coupling interactions between coils, shields, bore, and body. The achievable compression can enable GPU programming to accelerate the integral equation's solution in the case of clinically relevant voxel resolutions, for simulation setups that involve both local and remote interactions between conductive surfaces and tissue. Since the compressed coupling matrices can be stored and re-used without repeating the assembly process, the proposed method could facilitate iterative coil optimization pipelines that include the scanner bore and an RF shield. Finally, the idea of domain decomposition in the hybrid\hyp VSIE method could be easily extended to wire integral equations methods \cite{monin2020hybrid} to facilitate simulations in the case of coils made with wires rather than flat conductors \cite{tavaf2021self}.

\bibliographystyle{IEEEtran}
\bibliography{IEEEabrv,References}

\end{document}